\def\CSL1{\mbox{CSL-1}}

\def\CDF{\mbox{OACDF}}

\documentclass[useAMS,usenatbib]{mn2e}

\usepackage{graphicx}

\usepackage{natbib}

\title[The peculiar object \CSL1]{\CSL1: chance projection effect or serendipitous discovery
of a gravitational lens induced by a cosmic string?}

\author[M. Sazhin et al.]
{M. Sazhin $^{1,2}$, G. Longo $^{3,4}$, M. Capaccioli $^{1,3}$, J.
M. Alcal\'a$^{1}$, R.Silvotti$^{1}$, G.Covone$^{4}$,
\newauthor O.Khovanskaya$^{2}$, M.Pavlov$^{1}$, M.Pannella$^{1}$,
M.Radovich$^{1}$, V.Testa$^{5}$\\
$^1$ INAF - Osservatorio Astronomico di Capodimonte, via
Moiariello 16, I-80131 Napoli, Italy\\
$^2$ Sternberg Astronomical Institute, Universitetsky pr., 13,
119992, Moscow, Russia\\
$^3$ Dipartimento di Scienze Fisiche, Univ. Federico II, Polo
delle Scienze e della Tecnologia, via Cinthia, 80126 Napoli, Italy\\
$^4$ INAF - Telescopio Nazionale Galileo, Roque de Los Muchachos,
Santa Cruz de La Palma, 38700--TF, Spain P.O. Box 565\\
$^5$ INAF - Osservatorio Astronomico di Monte Porzio, Monte Porzio
Catone (Roma) Italy}

\pagerange{\pageref{firstpage}--\pageref{lastpage}} \pubyear{2002}

\def\LaTeX{L\kern-.36em\raise.3ex\hbox{a}\kern-.15em

    T\kern-.1667em\lower.7ex\hbox{E}\kern-.125emX}

\label{firstpage}
\begin{document}
\maketitle
\begin{abstract}
\CSL1\ (Capodimonte--Sternberg--Lens Candidate, No.1) is an
extragalactic double source detected in the \CDF\ ({\it
Osservatorio Astronomico di Capodimonte - Deep Field}). It can be
interpreted either as the chance alignment of two identical
galaxies at $z=0.46$ or as the first case of gravitational lensing
by a cosmic string. Extensive modeling shows in fact that cosmic
strings are the only type of lens which (at least at low angular
resolution) can produce undistorted double images of a background
source. We propose an {\em experimentum crucis} to disentangle
between these two possible explanations. If the lensing by a
cosmic string should be confirmed, it would provide the first
measurements of energy scale of symmetry breaking and of the
energy scale of Grand Unified Theory (GUT).
\end{abstract}
\begin{keywords}
cosmic strings; galaxies: general; cosmology: gravitational
lensing
\end{keywords}

\section{Introduction}
The Capodimonte - Sternberg - Lens candidate n.1 (or \CSL1) is a
peculiar object discovered in the \CDF\ ({\it Osservatorio
Astronomico di Capodimonte - Deep Field}): a medium-deep
multicolour imaging survey covering in three broad and in several
intermediate-width bands a $0.5\times1$ deg wide field located at
high galactic latitude. The \CDF\ data were collected at La Silla,
Chile, using the Wide--Field Imager at the ESO-MPI 2.2m telescope.
An additional set of frames in the $H$ band filter were taken in
March 2002 at the Telescopio Nazionale Galileo (TNG) with the Near
Infrared Camera and Spectrograph (NICS). In Table \ref{table1} we
summarize the most relevant information on the \CDF . More details
on the \CDF\ data reduction and calibration can be found in
Capaccioli et al. \cite{cap01}, and Alcal\'a et al. \cite{eso02}.

\begin{table}
\begin{center}
\begin{tabular}{|lccr|}
\hline
\hline
Filter & Exp. Time & PSF   & Phot. Err. \\
       & [hours]   & [arcsec]&            \\
(1)    & (2)       &  (3)  & (4)        \\
\hline
B      & 2.0       &  1.14  &      $\pm 0.11$      \\
V      & 1.7       &  1.01 &      $\pm 0.13$      \\
R      & 3.3       &  0.98 &      $\pm 0.21$      \\
H$^{*}$& 0.33      &  0.85 &      $ >0.2$     \\
\hline
753 nm & 6.5       &  0.87 &  $\pm 0.11$          \\
770 nm & 6.0       &  0.86 &  $\pm 0.12$          \\
791 nm & 6.5       &  0.97 &  $\pm 0.12$          \\
914 nm & 5.6       &  0.79 &  $\pm 0.13$          \\
\hline
\end{tabular}

* The H band covers only the \CSL1 region
\end{center}
\caption{Characteristics of \CDF. Column 1: photometric broad
bands and effective wavelenghts for narrow bands; column 2: total
exposure time of the final co-added image [in hours]; column 3:
FWHM of the resulting PSF [in seconds of arc] measured in
proximity of \CSL1; column 4: relative photometric accuracy
measured at the completion limit.} \label{table1}
\end{table}

The photometric criteria listed in Schneider et al. \cite{sch92}
as rules of thumb for identifying gravitational lens candidates,
i. e. the presence of at least two images with small (few arcsecs)
angular separation and having the same flux ratio in different
spectral bands, were applied to the \CDF\ first visually and then
on the matched catalogues, leading to the identification of
several faint gravitational lens candidates. After the rejection
of a few spurious objects, the list of candidates was reduced to
four objects (named by us Capodimonte-Sternberg Lens candidates or
\CSL1, \mbox{CSL-2}, etc ...).

\begin{figure*}
\begin{center}
\includegraphics[width=16.0cm]{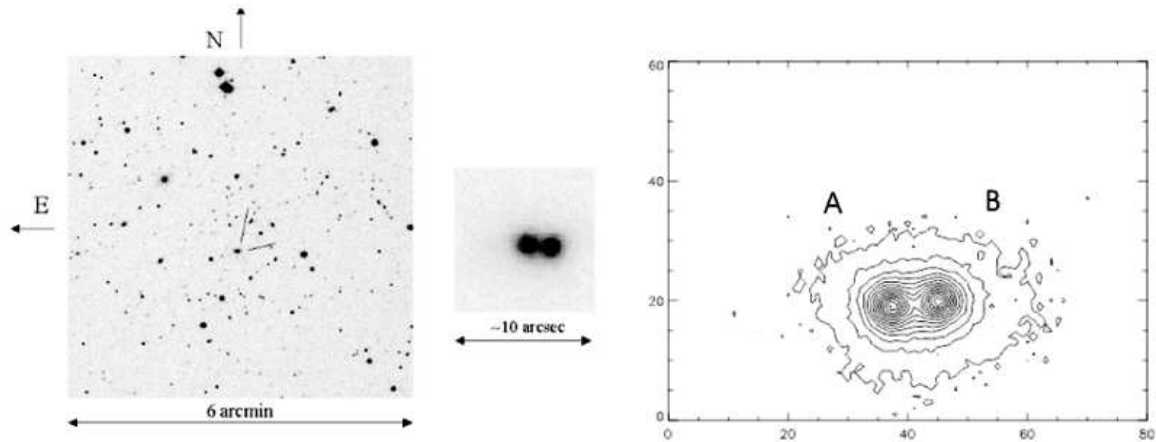}
\end{center}
\caption{Left panel and central inset: appearance of \CSL1\ in the
R band. Right panel: 2D contours of \CSL1\  from the near IR
($\lambda 914$) image. Coordinates are in pixels (1 px $=
0''.238$) and the two components are labeled A and B as in the
text. } \label{fig1}
\end{figure*}

So far, the spectroscopic follow-up's needed to confirm or
disprove the gravitationally lensed nature of the objects could be
performed only for the first candidate (\CSL1), which turned out
to be a rather interesting case.

\begin{table*}
\begin{center}
\begin{tabular}{|lcccrrrr| }
\hline
Band       & FWHM A & FWHM B & FWHM PSF& mag A            & mag B            &  $r_{e}^{A}$ & $\frac{r_{e}^{A}}{r_{e}^{A}}$  \\
           &[arcsec]&[arcsec]&[arcsec] &                  &
            & [arcsec]&\\
\hline
B          & 1.59   & 1.67   & 1.14    &  22.73$\pm .15$  & 22.57$\pm .15$   &              &              \\
V          & 1.59   & 1.67   & 1.01    &  20.95$\pm .13$  & 21.05$\pm .13$   &   6.3        & 1.4         \\
R          & 1.98   & 1.98   & 0.98    &  19.67$\pm .20$  & 19.66$\pm .20$   &   3.0        & 2.5         \\
H          & 1.19   & 1.11   & 0.85    &                  &                  &              &              \\
A753       & 1.11   & 1.19   & 0.87    &                  &                  &              &              \\
A770       & 1.27   & 1.27   & 0.86    &                  &                  &   7.4        & 0.6         \\
A791       & 1.67   & 1.59   & 0.97    &                  &                  &              &              \\
A914       & 1.27   & 1.27   & 0.79    &                  &                  &   8.8        & 1.4         \\
 \hline
\end{tabular}
\end{center}
\caption{FHWM, magnitudes and effective radii for the two
components of \CSL1\ . Column 1: photometric band; column 2 and 3:
FWHM size of component A and B, respectively; column 4: FWHM of
the PSF measured in a region close to \CSL1 ; column 5 and 6:
integrated magnitudes of components A and B respectively. These
values are provided only for the bands where accurate absolute
photometry could be performed. Column 7 and 8: effective radius
for the A component and ratio of the effective radii for the two
components, respectively. These values are provided only for those
bands where the profile was extended enough to allow a reliable
fit to a de Vaucouleurs law. } \label{table3}
\end{table*}

This paper is structured as follows: in Section 2 we summarize the
measured photometric and spectroscopic properties of \CSL1\ and in
Section 3 we discuss the possible explanations of the observables.
In Section 4, after introducing some aspects of the string
phenomenology, we present our model for the lensing by a cosmic
string and  the simulation performed in order to assess whether it
may explain the strange properties of \CSL1. Finally, in Section 5
we sumarise our results and discuss possible future observations.

\section{The observed properties of \CSL1\ }

\subsection{Morphological and photometric properties}

\CSL1\  consists of two sources (A and B, as marked in Fig.
\ref{fig1}) separated by 1.9 arcsec. Visual inspection shows that
in all bands, the two sources have (within the errors) the same
morphology: a bright nucleus surrounded by a faint halo with
undistorted and almost circular isophotes.

In Table \ref{table3} we list, in each of the \CDF\ broad bands,
the measured FHWM of the two components together with the FWHM of
the PSF measured on non--saturated stars near the position of
\CSL1\ . Also in Table \ref{table3} we give the integrated
magnitudes of the two components in the various bands. It is
apparent that, within the errors, the colors of the two components
are identical.

\begin{figure}
\begin{center}
\includegraphics[width=8.0cm]{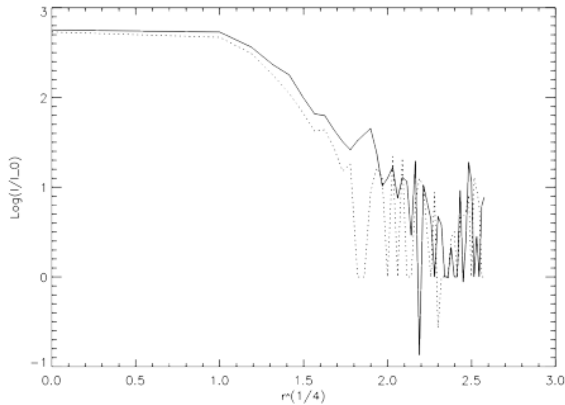}
\end{center}
\caption{Surface brightness profiles obtained for the the
components A (dashed line) and B (solid line) in the 914 \AA \
band. The profiles are normalized to the peak intensity and
plotted in $r^{1/4}$ units.} \label{fig2}
\end{figure}

In order to investigate the light profile, we performed a two
dimensional fit of the observed light distribution with a 2-D
Sersic profile $I_{S}(r) = I_{0} exp(-b(\frac{r}{r_c})^{1/n})$
convolved with the measured PSF, where:

$$r^2=\frac{1+e}{2}(x^2+y^2)+\frac{1-e}{2}(x^2+y^2)\cos 2\psi + \frac{1-e}{2}xy
\sin 2\psi $$

\noindent $x$ and $y$ being the Cartesian coordinates measured
from the central peak, $e$ the ellipticity of the corresponding
isophote, and $\psi$ the position angle of the isophote. The best
fit was obtained for $n=4$, id est for a ``de Vaucouleurs''
\cite{devauc} law. In Fig. \ref{fig2} we show the surface
brightness profiles obtained for the two components in the $914 \
\AA$ band; in ordinate we have the logarithm of the surface
brightness normalized to the peak value, while the abscissa is in
$pixel^{1/4}$. The slopes of the linear parts of the two profiles
are identical within the errors. For those bands where the surface
brightness profile covered a large enough range in magnitudes we
also attempted to derive estimates of the effective radii of the
two components (see \ref{table3}). These values, however, suffer
of very large errors due to the intrinsic faintness of the source
and to the inadequate sampling attained in some bands, and
therefore must be regarded as indicative only.

\begin{figure*}
\begin{center}
\includegraphics[width=14.8cm]{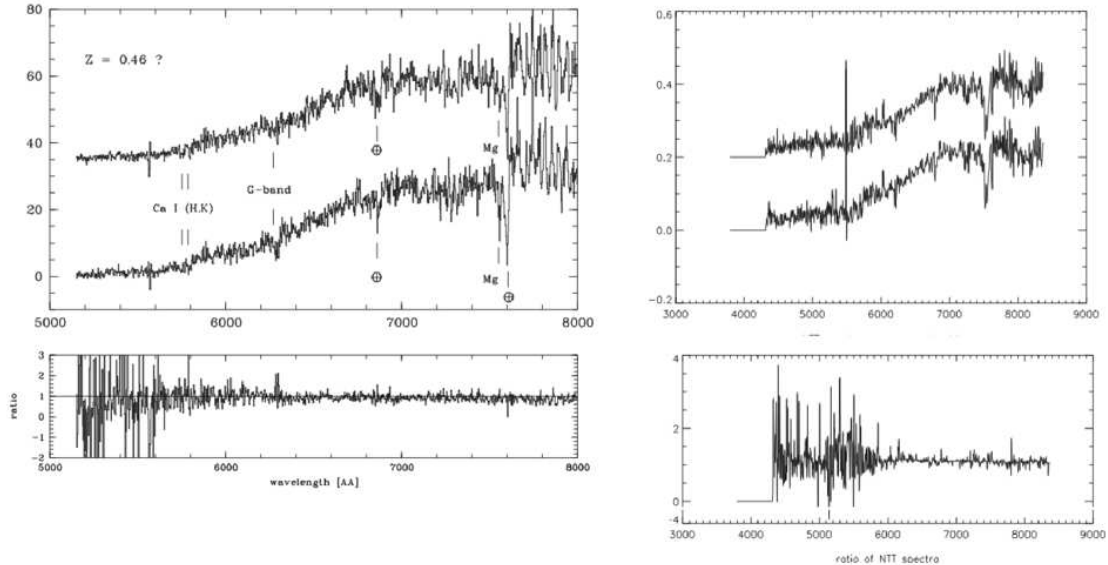}
\end{center}
\caption{Left panel: TNG spectra of the components of \CSL1 .
Right panel: NTT spectra. A vertical shift was introduced for
visualization purposes only. Lower panels: corresponding ratios
obtained by dividing the spectra of the two components.}
\label{fig3}
\end{figure*}

The two dimensional fit was also used to produce a residuals map
$(observed - model)$ of the light distribution which did not show
any systematic trend, with the possible exception of a very faint
light excess on the East side. We wish to stress, however, that
even in our deepest images ($R_{lim}\simeq 24.5$) we could not
detect any residual light in between the two images (see Fig.
\ref{fig9}).

Finally, we wish to emphasize that the fainter isophotes of the
two components, even though very noisy, have, within the errors,
almost identical shapes.

\begin{table}
\begin{center}
\begin{tabular}{|l|c|c|c|}

\hline
Equipment  & Exp.       &Spect.   &  Spectral  \\
           & time (hours) &resol.   &  range  \AA   \\
\hline
TNG + Dolores & 1  & 12 $\AA$  &  5200 - 7600  \\
NTT + EMMI    & 2  & 9.4$\AA$  &  4000 - 8500  \\
\hline
\end{tabular}
\end{center}
\caption{Spectroscopic material. Column 1: telescope and
spectrograph; column 2 total exposure time [in hours]; column 3:
spectral resolution measured by the FWHM; column 4: covered
spectral range.} \label{table2}
\end{table}

\subsection{Spectral properties}
We obtained two sets of long-slit spectra (with the slit aligned
along the direction joining the two images) of both components of
\CSL1 . The first one was obtained at the Telescopio Nazionale
Galileo (TNG) under non photometric conditions and therefore could
not be calibrated, while the second one was obtained at the
ESO-New Technology Telescope (NTT) in photometric conditions (cf.
Table \ref{table2}) and the spectrophotometric standards Hiltner
600 and Eg274 were also observed. Spectra were reduced using
standard IRAF routines and the sky was removed using the AFI
package \cite{Lor1993}.

In Figure \ref{fig3} we show both sets of spectra after applying -
for visualization purposes only - a slight vertical shift to the
spectrum of component A with respect to the spectrum of component
B. In the figure we also mark the position of the most significant
features which could be identified. No emission lines could be
detected in any of the available spectra.

The cross correlation between the spectra of the two components in
the same exposure, gives a differential radial velocity of $27\pm
25$ km/s, and the redshift is $z=0.46\pm 0.008$ (independently
from both data sets).

\begin{figure}
\includegraphics[width=8.0cm]{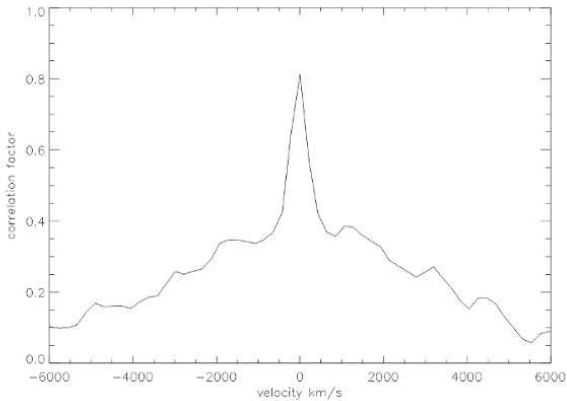}
\caption{Correlation coefficient of NTT spectra of the two
components of \CSL1\  with background profile removed.}
\label{fig4}
\end{figure}

The small difference which is observable in the vicinity of Ca
line, is very likely instrumental. In order to show the
correspondence between the two spectra, in Figure \ref{fig3} we
also show the ratio of the spectra of the two components, which,
within the errors is equal to 1 over the whole spectral range. In
Fig.\ref{fig4} we show the correlation function of the two
spectra, which confirms that the spectra are identical with a
confidence level higher than $99.9\%$.

\subsection{Putting all together}
The available photometric and spectroscopic data suggest that both
components of \CSL1\ are early-type galaxies at a redshift $z =
0.46$. We wish to stress that at such a distance (with $H_0=65$
km/s/Mpc, the distance is $\sim 1.9$ Gpc), the lower limit for the
linear separation between the two objects would be $\sim 20$ Kpc.
In Table \ref{table3} we give the absolute magnitude in each band
and the estimate effective radii for component A (component B has
identical values). All these facts converge in identifying the two
images of \CSL1\ as giant elliptical galaxies.

Whether the two components of \CSL1\ are two identical ellipticals
which happen to lay almost along the same line of sight, or rather
two images of the same background object produced by a
gravitational lens is much more difficult to disentangle. Here we
just want to emphasize that the lack of any excess light in
between the two images, implies that, if \CSL1\ is a lensed
object, then the lens is invisible and should be a very peculiar
``dark'' one \cite{Jac1998} which, furthermore, does not distort
the overall morphology of the lensed object.

As we shall demonstrate in Section 5, the only gravitational lens
able to produce such a morphology is a cosmic string but, before
examining this odd possibility, we need to rule out or at least
investigate all other possible, even though equally unlikely,
alternatives.

\subsection{\CSL1\ as a peculiarly obscured object}
The observed morphology could be the result of obscuration by a
gigantic strong dust lane. There are several reasons which allow,
however, to reject this hypothesis. First of all, with a minimum
separation between the two sources of 20 Kpc, it would be by far
the largest such structure ever observed; second, in order to
produce identical and symmetric sources it should have a perfectly
tailored shape. Furthermore, even in such unlikely case, such an
obscuring layer would fail to reproduce the light profiles
observed at the various wavelengths.

\begin{figure}
\begin{center}
\includegraphics[width=8.0cm]{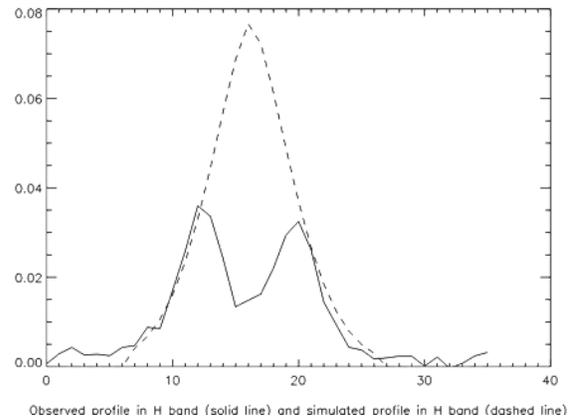}
\end{center}
\caption{Solid line: observed H band profile for \CSL1\ ; dashed
line: profile expected in the H band following the procedure
described in the text (dust index n=1).} \label{fig5}
\end{figure}

To prove it quantitatively we performed a simple simulation. The
underlying hypothetical galaxy was modeled assuming, as previously
done, a de Vaucouleurs profile and for the dust lane a standard
absorbing law given by $exp(-\tau (x))$, where $x$ is the
coordinate along the profile and $\tau$ is the geometry factor
describing the distribution of dust at a given $x$:
$\tau(x)=\frac{f(x)}{\lambda^n}$. In this formula, $n$ is the so
called dust index (cf. Hildebrand \cite{hil83}; Ferrari
\cite{fer02}; Chini \cite{chi84}) which, from optical to radio,
assumes values inside the range $1 \div 2 $. The above formulae
allow us to derive the light profile to be expected in any given
band relative to the $R$ frame assumed as template. In Figure
\ref{fig5} we show the expected and the observed H profiles for a
dust lane capable to reproduce the dip observed in the R profile.

As an additional test, in order to investigate whether the
combined effects of dust extinction and very strong colour
gradients could reproduce the observed morphology, we measured
colour in the wings and between peaks of the B, V, and R profiles.
The results show that colour gradients, if present, are smaller
than $0.1$ mag and therefore are of no help in explaining the
observed morphology by means of a dust lane.

Therefore we are left with only two possible explanations for
\CSL1 . Either we are dealing with i) the projection of two giant
elliptical galaxies which are identical (at a 99\% level of
confidence) in terms of magnitudes, colours, morphology and, what
is more relevant, also in terms of spectral properties, or ii) we
are seeing the effects of an unconventional and so far never
observed gravitational lens which does not distort extended
images.

\section{\CSL1\ as a gravitational lens}
\subsection{Compact lens models}

The peculiar properties of CSL-1 cannot be explained in terms of
lensing by a compact lens such as, for instance, the models listed
in the Keeton catalogue of models (Keeton 2002). As a template
case, let us take into consideration that of a Singular Isothermal
Sphere (SIS), id est the most commonly used (and reliable) model
for a gravitational lens. In Fig.6 (left panel) we show the
results of the simulation of a SIS lensing effect for an extended
object: as it can be easily seen, the SIS model produces a
distortion of the outer isophotes. The simulation was performed as
follows: we assumed a "de Vaucouleur"-like 2D light distribution
and then mapped it to the lens plane accordingly to the
gravitational field generated by a SIS. Afterwards, the modelled
image was smoothed with the observed Point Spread Function and a
realistic (id est with the same statistics of the real image)
noise was added. As optimization value we took the difference
between the observed and modelled images squared and optimized
over the parameters of the light distribution and of the SIS. The
distorsions introduced by the SIS are clearly visible. In Fig.6
(right panel) we show the residuals between the modelled and
observed images: as it can be seen, the residuals exceed by
5$\sigma$ the noise level. This behavior of the residuals is
confirmed also by the analysis performed in all other available
bands and we wish to stress that, since in different bands the
noise in uncorrelated, the residuals may be considered as
independent events and, therefore, the composite probability of
having the same fluctuations in the same positions is absolutely
negligible.

Furthermore, in the unlikely case of ordinary lensing by a massive
object, one could also estimate some characteristics of the lens.
First of all we can restrict the redshift range for the lens
inside $0.1 < z < 0.4$. The velocity dispersion (mass parameter
for a SIS lens) should therefore be in the range 200 km/s up to
400 km/s. Such value is typical of a giant galaxy and therefore,
if we assume a lens made of ordinary matter, it should be clearly
visible in our images which is not a case. Therefore one have to
suppose that it is a case of dark lens. As stressed above, such a
''dark'' gravitational lens, while being completely invisible,
would introduce significant and measurable distortions in the
lensed images (producing arc-like features) which are not observed
in CSL-1.

\begin{figure*}
\begin{center}
\includegraphics[width=14.0cm]{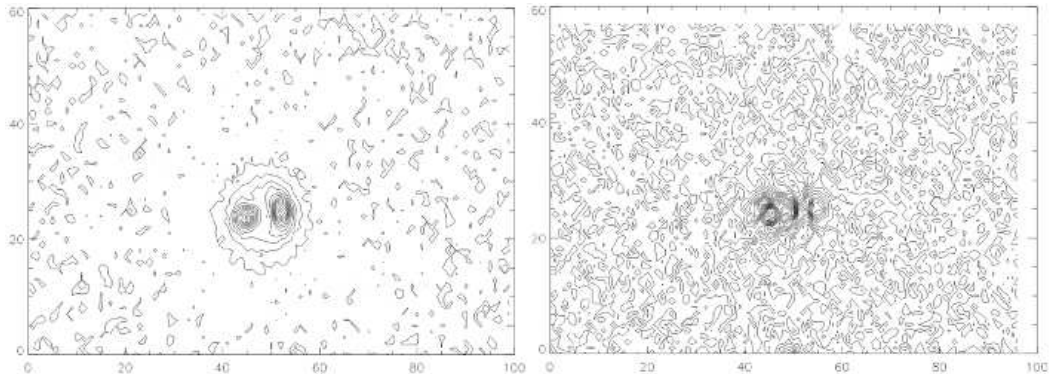}
\end{center}
\caption{Left: contours of the primary and secondary images in the
case of an optimal SIS lens in the band 914 $\AA$. The images are
convolved with PSF of the band. The distortion of the outer
isophotes is clearly visible. Right: residuals (modelled -
observed) for the 914 $\AA$ image. } \label{fig6}
\end{figure*}

\subsection{Cosmic string model}
Cosmic strings were introduced in theoretical cosmology by Kibble
\cite{kib76}, Zeldovich \cite{zel80}, Vilenkin \cite{vil81}, Gott
\cite{got85}, de Laix et al. \cite{lai96}, Bernardeau et al.
\cite{ber01}. Vilenkin \cite{vil81} also suggested that they could
be revealed by means of gravitational lensing phenomena.

From a physical point of view, a cosmic string is characterized by
a mass per unit length $\mu$ and by a tension $T=\mu$ along the
string. Therefore, a string has only one parameter defining both
its microscopic physical properties and the macroscopic properties
of the surrounding space-time. The microscopic characteristic of
the string is the diameter, which is of the order of the square
root of $\mu^{-1}$. Below we shall use Planck units, so that $\mu
=1$ corresponds in usual units to $1.4\cdot 10^{28}$ g/cm and the
diameter of a cosmic string would be of the order of $10^{-33}$
cm. All modern theories, however, predict $\mu \ll 1$ thus
implying a larger but still microscopic value for the expected
diameter of a string. We can therefore safely assume that, on a
cosmic scale, the diameter of the string can be neglected with
respect to all the other linear quantities involved.

From the lensing point of view, a cosmic string produces a conical
space time where the cusp of the cone coincides with the position
of the string. The complete turn in a plan perpendicular to the
string gives the total angle $\Phi$ (smaller than $2\pi$). The
difference $2\pi - \Phi$ is called ``angle deficit'' and it is
defined by the string density:

\begin{equation}
D=8 \pi \mu \label{def}
\end{equation}

\noindent in the case of $\mu \ll 1$. In order to understand the
lensing properties of a cosmic string let us assume the following
geometry: a distant point-like source (for instance a QSO) and a
cosmic string interposed along the line of sight and at rest with
respect to the observer. Let the angular cosmological distance
from the observer to the string be $R_s$ and that between the QSO
and the observer $R_q$. If the angular distance between the QSO
and the string (as seen from the observer) is less than $D$, then
the observer sees two images of the QSO separated by the angle
(Gott, 1985):

\begin{equation}
\Delta \theta = D \sin \alpha \frac{R_q-R_s}{R_q}, \label{sep}
\end{equation}

where $\alpha$ is the angle between the string and the vector
coinciding with the observer and the point source.

In what follows we shall assume for simplicity that $R_s \ll R_q$
and $\alpha =\pi/2$  that  is $\Delta \theta = D$. The proper
distances depend on the cosmological model (we shall assume a
standard cosmological model: FRW metric with flat space
$\Omega_{tot} = 1$, and $\Omega_{\Lambda}= 0.7$, $\Omega_{CDM} =
0.3$ according to modern measurements de Bernardis et al.,
\cite{dber01}). When the source is still far away from the string,
no lensing is induced and the image is unperturbed, while when the
angular separation reaches $D$, the lensing phenomenology sets in
and a second image of the source begins to form on the opposite
side of the string, at an angular distance $D$ from the source
position. We wish to stress that, on the contrary of what happens
in other lensing models, the lensing by a cosmic string does not
introduce any amplification of the signal from the background
source.

\begin{figure*}
\begin{center}
\includegraphics[width=12.0cm]{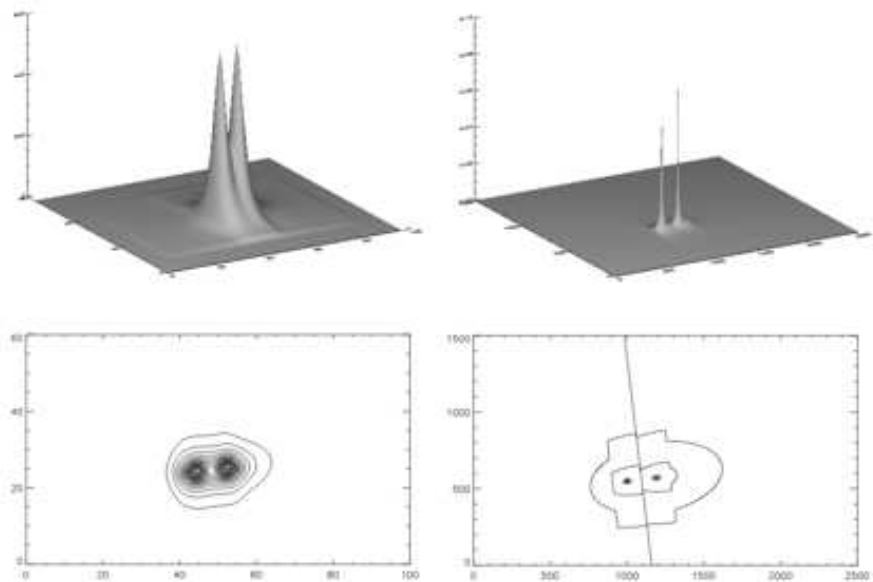}
\end{center}
\caption{Upper left and lower left panels: 3-D light distribution
and isophotal contours obtained after convolving our model with
the observed PSF. Upper right and lower right panels: the same but
convolving for an HST like PSF with a FWHM of 40 milliarcsec. }
\label{fig7}
\end{figure*}

\begin{figure}
\begin{center}
\includegraphics[width=8.0cm]{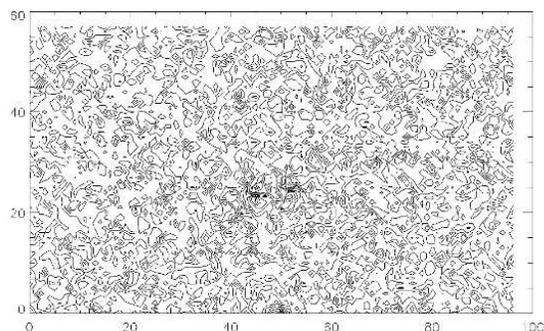}
\end{center}
\caption{Contours of the residuals (observed - modelled) for the
light distribution of \CSL1\  (in the 914 nm band). The maximum
value of the residual is $2\sigma$ above the background.}
\label{fig9}
\end{figure}

In Figure \ref{fig7} we show (as 3-D light distribution and as
isocontours) the results of our simulation of the observed
photometric properties of \CSL1\ in the band $\lambda = 914 \
\AA$. As before, the background source was modelled using de
Vaucouleurs law. The figure gives the outcome of the simulation
for two different seeing: the observed one and a 40 milliarcsec
seeing assumed to be representative of the HST average operating
conditions. Finally, in Figure \ref{fig9} we plot the isocontours
for the residuals (observed - modelled) of our best fitting model.
No systematic trend can be observed. As it can be seen, lensing by
a cosmic string should produce a characteristic signature
consisting in isophotes with very sharp edges. This signature,
however cannot be observed under average seeing conditions. In
fact, if the seeing FWHM is comparable with the size of the
secondary image (as it is the case for our data), the observer
sees smooth and nearly circular images with the secondary one
slightly fainter.

\section{Discussion and conclusions}
As discussed above, the optical morphology of the candidate lens
\CSL1\  can be explained in terms of lensing by a cosmic string.
Even though this may seem unlikely due to the expected rarity of
such strings and to the fact that on the basis of present
observational evidence we cannot yet rule out the only other (also
unlikely) possible explanation in terms of projection effects, we
want to stress that, if confirmed, this would be the first clear
cut evidence ever for a cosmic string, and would therefore have a
far reaching impact in both cosmology and particle physics. There
is, however, an easy way to disentangle between the two possible
explanations. As shown above, a cosmic string would introduce a
clear cut signature in the shape of the lensed object which can be
detected in high angular resolution optical data. In the case of
positive detection, the parameters derived for the cosmic string
will allow us to determine its linear density and thus measure the
energy scale of simmetry breaking and of the energy scale of Grand
Unification Theory (GUT).

Observation in radiofrequency bands would also be needed. A
straight long string induces temperature fluctuation on the cosmic
microwave background radiation, and is therefore expected to
produce a linelike discontinuity at the place of the string. The
amplitude of fluctuation is given by Kaiser and Stebbins
\cite{kai84}:
$$
\frac{\delta T}{T} \approx 8\pi \mu v
$$
\noindent here $v$ is the velocity of string perpendicular to the
line of sight in the units of light velocity.  Usually the
velocity of cosmic string is expected to be close to unity due to
stresses which are inside the string. If the velocity is close to
unity, one can expect significant anisotropy which will form a
strip of $\delta T/T$ along the position of the string with
angular width of the order of 2" and amplitude $\sim$~30~$\mu$K.
This strip would cross the optical images of CSL-1 perpendicularly
to the line connecting the two components. Although the expected
angular scale is very small, the amplitude is easily detectable
with modern radiometers (see, for instance, de Bernardis et al.
\cite{dber01}).

If we accept the interpretation  of \CSL1\ as a gravitational lens
produced by a cosmic string, it is possible to derive the scale of
energy at which the symmetry breaking occurred. The physical
mechanism which produces cosmic strings is related to the phase
transition in the early Universe: cf. Gott, \cite{got85} and
Dolgov Sazhin and Zeldovich, \cite{dol90}. This phase transition
took place when the temperature of the Universe felt below a
critical temperature defined by some energy scale which, on the
other end, defines the unification of physical interactions, and,
probably, also the main parameters of the inflation. The distance
between the peaks of the two images $\sim$~2" of \CSL1 roughly
corresponds to the "deficit angle" $D$. One can therefore estimate
the density of the string as $4\cdot 10^{-9}$ $m_{pl}^2$, or in
CGS physical units $5.4 \cdot 10^{21}$ g/cm. This value, if our
interpretation of \CSL1 is correct, may be used to estimate many
physical parameters.

For grand unified strings one can expect
\begin{equation}
\mu \sim \frac{m^2}{\alpha} \label{mu}
\end{equation}

\noindent where $\alpha$ is a running coupling constant and $m$ is
the typical mass scale of the symmetry breaking. Particle
physicists believe that symmetry breaking took place at an energy
scale of the order of $E \sim 10^{16 \pm 0.3}$ GeV. This value is
derived from the extrapolation of accelerators data
Klapdor-Kleingrothaus, and Staud, \cite{kla91}. Near this scale,
all running coupling constants of physical interactions merge, and
the value of united coupling constant is:
\begin{equation}
\alpha^{-1} = 25.7 \pm 1.7
\label{mu1}
\end{equation}

Taking into account (\ref{mu}) and (\ref{mu1}) and the value of
the deficit angle derived from our data, one can estimate the mass
scale of symmetry breaking as $2\cdot 10^{15}$ GeV, which is in
reasonable agreement with extrapolated accelerators data.

\section*{Acknowledgements}
M.V. Sazhin acknowledges the Capodimonte Astronomical Observatory
for hospitality and financial support, also acknowledges the
financial support of RFFI grant 00-02-16350. This research was
partly funded by the Italian Ministry for Public Instruction,
University and Research through a COFIN grant. Authors are also
indebted to G. Busarello for fruitful discussions and to the ESO
and TNG staff for support during the observations.

\end{document}